# Ferroelectric Domain Morphology Evolution and Octahedral Tilting in Lead-Free $(Bi_{1/2}Na_{1/2})TiO_3$-$(Bi_{1/2}K_{1/2})TiO_3$-$(Bi_{1/2}Li_{1/2})TiO_3$-$BaTiO_3$ Ceramics at Different Temperatures

Cheuk Wai Tai[*†], Siu Hong Choy and Helen L. W. Chan

Department of Applied Physics and Materials Research Centre, The Hong Kong Polytechnic University, Hung Hom, Kowloon, Hong Kong, People's Republic of China

**Abstract**

Observations by *in-situ* transmission electron microscopy (TEM) reveal the changes in domains morphology in $0.885(Bi_{1/2}Na_{1/2})TiO_3$-$0.05(Bi_{1/2}K_{1/2})TiO_3$-$0.015(Bi_{1/2}Li_{1/2})TiO_3$-$0.05BaTiO_3$ ceramics. Evolution of the macroscopic ferroelectric domains has been observed at 25 °C to 180 °C. The ferroelectric domains and their walls reduce progressively with increasing temperature. In the selected-area electron diffraction (SAED) patterns the crystallographic evidence of antiferroelectric domain is not found over the temperature range, which provides the evidence that the ferroelectric-antiferroelectric transition does not occur in the ceramic. The temperature-dependent forbidden 3/2 1/2 0 and superstructure 3/2 1/2 1/2 reflections indicate the presence of an in-phase and anti-phase octahedral tilting in the ceramic, respectively. Dark-field TEM images show that the size of the octahedral tilted domains is smaller than 30 nm across and these domains are uniformly distributed in the grains. The influence of the localized octahedral titling on the stability of ferroelectric domains and the temperature-dependent properties is discussed in this paper.

[*]Author to whom correspondence should be addressed. Email: cheukw.tai@gmail.com
[†]Present address: Department of Materials and Environmental Chemistry, Stockholm University, S-109 61, Stockholm, Sweden.

## I. Introduction

Developments of lead-free piezoelectric ceramics have drawn great attention as a technological issue since the restriction of the use of certain hazardous substances (RoHS) and the regulations on waste electrical and electronic equipment (WEEE) have been proposed. Most of the studies are focused on perovskite $Bi_{1/2}Na_{1/2}TiO_3$ (BNT)-based materials.[1,2,3,4,5,6,7,8] $Bi_{1/2}Na_{1/2}TiO_3$-$BaTiO_3$ (BNT-BT)[5] and $Bi_{1/2}Na_{1/2}TiO_3$-$Bi_{1/2}K_{1/2}TiO_3$ (BNT-BKT)[6] solid solutions are of particular interest because of their outstanding dielectric and piezoelectric properties at the compositions near the morphotropic phase boundary analogous to the $PbTiO_3$-based solid solutions. By mixing these two solid solutions, a new ternary system, $Bi_{1/2}Na_{1/2}TiO_3$-$Bi_{1/2}K_{1/2}TiO_3$-$BaTiO_3$ (BNT-BKT-BT), is obtained.[7,8,9,10] It is found that the dielectric and piezoelectric properties in some compositions of BNT-BKT-BT are comparable to $Pb(Zr,Ti)O_3$-type materials.[11,12]

The BNT-based ceramics have a high Curie temperature ($T_c$) of about 300 °C. However, these ceramics undergo another phase transition below $T_c$ that is known as depolarization temperature ($T_d$), which often occurs below 200 °C. When the depolarization takes place at this phase transition, the piezoelectric properties of the BNT-based ceramics are reduced significantly or even vanished. In a practical point of view, $T_d$ is a more important parameter than $T_c$ because it is the operational limit of a BNT-based device. At the moment, the mechanism of depolarization is still not clear. The generally accepted explanation is that the depolarization is caused by a ferroelectric-antiferroelectric transition, which is claimed by the observation of the peculiar propeller-shaped hysteresis loops (also called double loops). The occurrence of the double loop, in fact, can be induced by various mechanisms. The most obvious one is the electric-field induced antiferroelectric-ferroelectric transition in an antiferroelectric material below $T_c$, for example in $PbZrO_3$.[13] The double loop can also be obtained at an electric-field induced transition occurred at several degrees higher than the first order paraelectric-ferroelectric transition in a ferroelectric material, as in $BaTiO_3$ and highly ordered $Pb(Sc_{1/2}Ta_{1/2})O_3$.[14,15] The double loops can also be induced by the behavior of ferroelectric domains and their walls. In aged ferroelectrics,[16,17] the double loop is originated from the contribution of the domain configuration with non-zero polarization after removing the high applied electric-field. The domain walls do not return to their original states. This effect is known as the constriction of the hysteresis loop. On the contrary, the constrained polarization state can occur in ferroelectrics without aging. The ferroelectric domains can recover their initial configurations when the electrical bias has been removed. But the defected dipoles, which are induced by dopants, vacancies or local strains, obstruct the reversal of the switched polarization.[18] The overall polarization response can give rise to the double loop. Moreover, the polarization reversal occurred in a ferroelectric solid solution, in which the paraelectric and ferroelectric phases co-exist, can give the double hysteresis loop, for example in $(Pb_xCa_{1-x})TiO_3$.[19] In the BNT-based ceramics, it is suggested that the double-loop behavior could be induced by the transformation of macroscopic to microscopic ferroelectric domains,[20] in addition to the antiferroelectric-ferroelectric transition.

Each of the above mechanisms can be distinguished by their characteristic of the hysteresis loop and the corresponding crystallographic features. Unfortunately, the crystallographic studies of BNT-based solid solutions are rarely reported.[21,22,23,24] *In-situ* transmission electron microscopy (TEM) can be used to trace the evolution of domain patterns with temperature. In addition to imaging the domain morphology, crystallographic details can be revealed by electron diffraction in a transmission electron microscope simultaneously. The commensurate antiferroelectric state in a perovskite-structured material is characterized by the presence of 1/4 {110} (pseudo-cubic) forbidden reflections in a selected-area electron diffraction (SAED) pattern.[25,26] On the other hand, when an incommensurate antiferroelectric state establishes in the material, satellite reflections along the <110>* directions or the 1/x {110} forbidden reflections can be observed in a SAED pattern.[15,27,28,29] The additional modulated structures given by either antiferroelectric state can also be seen in the high-resolution TEM images.[26,29]

In this paper, we report the results of the *in-situ* TEM studies on $0.885(Bi_{1/2}Na_{1/2})TiO_3$-$0.05(Bi_{1/2}K_{1/2})TiO_3$-$0.015(Bi_{1/2}Li_{1/2})TiO_3$-$0.05BaTiO_3$ (abbreviated as BNKLBT-1.5). The composition of the BNT-BKT-BT ceramic without doping $Li^+$ is close to the morphotropic phase boundary.[9] By doping with 1.5 mole % $Li^+$, the piezoelectric and dielectric properties of the ceramics are improved significantly and are comparable to certain PZT-based materials.[30,31] The study of ferroelectric domain morphology and the search for the crystallographic evidence of antiferroelectric domains are carried out at various temperatures.

**II. Experimental**

BNKLBT-1.5 ceramics were fabricated by conventional mixed oxide route.[30,31] The precursor materials were synthesized using commercially available bismuth (III) oxide ($Bi_2O_3$, 99.9%, Acros Organics), sodium (I) carbonate ($Na_2CO_3$, 99.5%, Wako Chem. Co. Ltd), potassium (I) carbonate ($KCO_3$, 99.0%, Panreac Quimica SA), lithium (I) carbonate ($Li_2CO_3$, 99+%, International Laboratory), barium (II) carbonate ($BaCO_3$, 99.9%, Wako Chem. Co. Ltd) and titanium (IV) oxide ($TiO_2$, 99.9%, Aldrich Chem. Co.). The amounts of $Bi_2O_3$, $Na_2CO_3$, $KCO_3$, $LiCO_3$, $BaCO_3$ and $TiO_2$ used follow the chemical formula. The precursors were mixed together by ball-milling in ethanol with zirconia media for 10 h. The calcination was performed at 800 °C for 2 h. The calcined powder was ball-milled again for 10 h. After the slurry was dried, polyvinyl alcohol (PVA) as a binder was added to the powder. Pellets were obtained by uniaxial pressing. The ceramics were submerged in the powder of the same composition and sintered at 1170 °C for 2 h in air.

Prior to the ferroelectric hysteresis loop measurements, silver paste was applied on both surfaces of the polished ceramics and fired at 650 °C as electrodes. The hysteresis loops of the ceramics were obtained by a conventional Sawyer–Tower circuit at 100 Hz. For the TEM study, electron-transparent samples were mechanically polished to a thickness of 20 μm prior to ion-milling at 3.5 kV (Gatan PIPS 691). The *in-situ* TEM studies were carried out on a JEOL-2011 microscope operated at 200 kV and equipped with a double-tilt heating stage (Gatan Model 652). The holding time at each temperature was over 30 minutes.

**III. Observations**

*(1) Ferroelectric hysteresis*

Ferroelectric hysteresis (D-E) loops of the BNKLBT-1.5 ceramic measured at 25 °C and 160 °C, respectively, are shown in Fig. 1. At 25 °C, an ordinary square-like hysteresis loop can be observed. The values of the remanent polarization ($P_r$) and coercive field ($E_c$) of the ceramic are 32.4 µC/cm$^2$ and 3.4 MV/m, respectively. In contrast, a peculiar propeller-shaped hysteresis loop was obtained at 160 °C, which is above $T_d$ (142 °C).[31] The values of $P_r$ and $E_c$ are 2.4 µC/cm$^2$ and 0.3 MV/m, respectively, which are decreased by an order of magnitude. This indicates that at 160 °C the spontaneous polarizations in the sample are reduced significantly. However, a saturation of polarizations can be seen and the maximum dielectric displacement ($D_{max}$) is also maintained as the value at 25 °C (32.4 µC/cm$^2$). These results show that at 160 °C the polarizations can be aligned to the saturated state, of which the configuration is similar to that at 25 °C, but a large portion of the polarizations cannot be sustained when the applied electric field was reduced to 1 MV/m.

*(2) Ferroelectric domains*

Fig. 2(a) shows the domain patterns recorded along [-110] at 25 °C. The main crystallographic directions are given, according to the corresponding SAED pattern of the same area shown in Fig. 2(b). Two kinds of domain configurations that are herringbone-like and parallel bands (indicated by white arrows) are observed. However, some small domains and their curved walls, which do not follow any crystallographic direction, are observed, indicating the presence of either 180° domains or other phases. In the herringbone-like configuration, the width of the domains is about 50 to 100 nm. The fringes are identified as the traces of (101) and (011) domain walls. These walls are inclined to the electron beam because no splitting along their corresponding directions is observed in the SAED pattern. The (101) and (011) domains walls meet at the (110) (white line) and (001) walls (black line). The angles between these two sets of domains are about 109° and 71° for the (110) and (001) walls, respectively. Some of these domains walls are terminated at two invisible boundaries which corresponds to the (-1-10) or (111) planes. These observations are similar to those in the rhombohedral phase PZT.[32] In the parallel band configuration, the domains point in [-1-1-1] and their walls can be identified as inclined and misoriented (101) walls. Bifurcations are also observed at the ends of the domains.

In a rhombohedral ferroelectric phase, the splitting of the reflections in a SAED pattern should be relatively small and it can be observed clearly as the higher order reflections. The direction of the splitting should also be perpendicular to a twin plane (or domain wall).[33] In Fig. 2(b), the splitting of the reflections is relatively large and also not perpendicular to one of the {110} or {100} domains walls. These do not seem to be the characteristic of a pure rhombohedral phase. In fact, it is found that the splitting is generated by a superposition of two SAED patterns with a rotation of ~0.9° that is similar to the TEM observations of domains in the PZT ceramics at MPB.[34] In the BNKLBT-1.5 ceramic, the coexistence of the rhombohedral and tetragonal ferroelectric phases is expected because the composition is at the MPB of the BNT-BKT-BT solid solutions.[30,31]

Fig. 2(c)-(f) show the domain patterns obtained at 100 °C, 140 °C, 160 °C and 180 °C, respectively. At 100 °C, some herringbone-like domains disappeared and a number of bifurcated domains at the tips of the parallel band domains were merged. The (001) domain wall, which is marked by a black line in Fig. 2(a), vanished. The associated [-1-11] domains extend to the region, which was formerly separated by the (001) wall at 25 °C. Both kinds of macroscopic domains further reduced when temperature increased to 140 °C which is close to $T_d$ (142 °C). Fig. 2(c) shows that the domains become shorter and narrower than those observed at 100 °C. The decrease of domain states leads to the reduction of polarization and consequently the piezoelectric properties. At 160 °C, some domains walls can still be seen although most of the macroscopic domains disappear. When temperature further increases to 180 °C, no herringbone-like domains can be observed but a few parallel band domains still exist. It should be noted that the small domains, which are not followed any predicted crystallographic directions, exist at all temperatures. Therefore, it is seen that at the low temperature, the ceramic contains both macroscopic and small

domains; whereas at the higher temperature, the macroscopic domains are reduced significantly but do not vanished completely. The macroscopic and small domains can be re-oriented by applying a high electric-field. However, at higher temperatures, they return to their original favorite states after removing the high applied field. The consequence is the propeller-shaped hysteresis loop with lower $P_r$ and $E_c$ (see Fig. 1). In addition, it is worthy to note that the domain transformation does not take place at $T_d$ immediately. In fact, the macroscopic domains are reduced progressively and continuously as temperature increases.

*(3) Selected-area electron diffraction*

Fig. 3(a)-(e) show the SAED patterns recorded along a <001> zone axis at different temperatures of a grain of BNKLBT-1.5 ceramic. The patterns are indexed according to space group Pm3m. In addition to the fundamental reflections, the forbidden reflections 1/2 1/2 0, also known as the γ reflections,[35] are observed in the pattern. The intensity of these forbidden reflections appears weak. One of the γ reflections 3/2 1/2 0 is marked by the white arrows. These forbidden reflections can be generated by either (i) the anti-parallel displacements of A-site cations along the <110> that can lead to the antiferroelectric state[35,36] or (ii) octahedral tilting occurred in the crystal.[37,38] The latter is the most likely cause of the development of the γ reflections in BNKLBT-1.5 ceramic.

In Fig. 3(a)-(e), it is found that the intensity of the γ reflections decreases with increasing temperature. The two proposed mechanisms to generate the γ reflections can show this temperature dependency because both of them involve structural change. However, we can distinguish the octahedral tilting from the A-site anti-parallel displacements by examining the intensity of the 1/2 1/2 0 reflections. Such forbidden reflections can be observed in Fig. 3(a) but they are weaker than the other higher-order γ reflections. Fig. 3(b) shows that the 1/2 1/2 0 reflections disappear when temperature has been increased to 100 °C. The lowest order of the γ reflections in the SAED pattern is the 3/2 1/2 0 reflections instead of the 1/2 1/2 0 reflections. The disappearance of the 1/2 1/2 0 reflection is also found in the SAED patterns recorded at 140 °C, 160 °C and 180 °C shown in Fig 3(c)-(e), respectively. If the anti-parallel displacements of the A-site cations establish in a perovskite crystal, the lowest order of the γ reflections should be the 1/2 1/2 0 reflections and their intensity is the strongest one. However, it is not the case in BNKLBT-1.5 ceramic. In addition, the evidence for commensurate and incommensurate antiferroelectric domains, which are the forbidden 1/4 1/4 0 and 1/x {110} reflections, respectively, are not found at the temperature range from 25 °C to 180 °C. It is confirmed that neither kind of antiferroelectric domains exists in BNKLBT-1.5 ceramic.

The presence of γ reflections indicates that the octahedral tilt system is the in-phase mode. In such system, the lowest order of the γ reflections is the 3/2 1/2 0 reflections.[39] It should be remarked that in the present study the SAED patterns were taken from the grains revealing ferroelectric domains. The appearance of the domains and walls implies that the TEM sample is not thin enough to avoid dynamical diffraction condition completely. The kinematically forbidden 1/2 1/2 0 reflections can be obtained by the double diffraction route: (3/2 1/2 0) + (-1 0 0). It can be seen that the intensity of the resultant 1/2 1/2 0 reflections strongly depends on the intensity of the 3/2 1/2 0. Therefore, we cannot observed the 1/2 1/2 0 reflections when the 3/2 1/2 0 reflections are weak, especially at high temperatures.

In the in-phase octahedral tilt systems, the γ reflections can be shown in some <001> and <111> zone axes SAED patterns but not in the <011> zone axes one.[35] Fig. 4 is a <111> zone axis SAED pattern recorded at 25 °C. The γ reflections and double diffracted reflections are observed in the figure. We would like to mention that in this <111> zone axis SAED pattern the temperature dependency of the γ reflections is similar to the observations shown in Fig. 3.

Fig. 5(a)-(e) shows the SAED patterns of a BNKLBT-1.5 grain obtained along a <011> zone axis at different temperatures. Fig. 5(a) obtained at 25 °C shows the superstructure 1/2 1/2 1/2 reflections. This kind of reflections is also known as the α reflections.[35] But no γ reflections can be observed in the <011> zone axes patterns. Accordingly, it is confirmed that the in-phase octahedral

tilt occurred in BNKLBT-1.5 ceramic. All the four in-phase octahedral tilt systems, $a^0a^0c^+$, $a^+a^+a^+$, $a^0b^+b^+$ and $a^+b^+c^+$, are possible.

It is similar to the case of the γ reflections that two mechanisms can generate the α reflections. In general, the 1/2 1/2 1/2 reflections indicate the 1:1 cation structural order existed in a complex perovskite.[40] On the other hand, the appearance of the α reflections can associate with the anti-phase octahedral titling in the sample. These two mechanisms can be differentiated by the stability of such reflections at different temperatures.

Fig. 5(b) is a SAED pattern taken from the same region as Fig. 5(a) but at 100 °C. The intensity of the α reflections decreases significantly. The higher the temperature, the weaker the α reflections are seen in Fig. 5(c)-(e). At 180 °C, the α reflections are vanishingly weak, in Fig. 5(e), indicating that the long-range 1:1 cation order does not exist in BNKLBT-1.5 ceramic. This result is consistent with the previous observations of BNT, BKT and their solids solutions, in which the short-range 1:1 cation order was found only.[21,41,42,43,44] In the present study, the thermal energy is definitely insufficient for stimulating the cation migration. We can deduce that the α reflections are not originated from 1:1 cation order.

According to the existence of the temperature-dependent α reflections, the anti-phase octahedral tilting occurs in BNKLBT-1.5 ceramic. Similar to the case of the γ reflections, the lowest order of the α reflections is 3/2 1/2 1/2 in an anti-phase octahedral tilt system.[35,39] The observed 1/2 1/2 1/2 reflections are the consequences of double diffraction. For example, the route can be (3/2 1/2 1/2) + (-1 0 0). In contrast to the SAED patterns along the <001> zone axes, some of the <011> zone axes patterns show no α reflections, as in Fig. 6. This helps to rule out some anti-phase tilt systems, besides indicating that the 1:1 cation order may only exist in atomic scale. In the $a^0a^0c^-$, $a^0b^-b^-$, $a^-a^-a^-$ and $a^-b^-b^-$ systems, the γ reflections disappear in certain <110> zone axes, which are perpendicular to the tilt axes. Therefore, the $a^0b^-c^-$ and $a^-b^-c^-$ tilt systems can be eliminated because they generate the γ reflections in all the twelve <110> zone axes SAED patterns. Unfortunately, the in-phase and anti-phase octahedral tilt systems occurred in the BNKLBT-1.5 ceramic cannot be determined because the disappearance of a particular row of γ and α reflections in the SAED patterns is not observed by tilting the sample systematically in the microscope. Although the $a^0a^0c^+$ and $a^-a^-a^-$ systems have been reported in the tetragonal and rhombohedral phases of BNT-BKT solid solutions, respectively,[21] other tilt systems may also establish in the lower symmetry phases of the BNKLBT-1.5,[23] which can be developed by the different combination of A-site cations.

We exclude all the four mixed octahedral tilt systems that are $a^-a^-c^+$, $a^+a^+c^-$, $a^0b^-c^+$ and $a^-b^-c^+$ although the in-phase and anti-phase tilt systems occur in BNKLBT-1.5 ceramic simultaneously. The mixed tilt system is indicated by the appearance of the additional forbidden reflections 1/2 1 1, besides the γ and α reflections.[38] However, such additional reflections have never been found in the SAED patterns. It is concluded that the mixed octahedral tilting does not exist in BNKLBT-1.5 ceramic as well as the anti-parallel displacements of the A-site cations along the <001>.

*(4) Dark-field TEM images of forbidden reflections*

In order to determine the size and distribution of the octahedral titled domains, dark-field TEM imaging was employed. Figs. 7(a) and (b) are the dark-field images of the octahedral titled domains at 25 °C. In Fig. 7(a), the in-phase titled domains were imaged using one of the α reflections along a <001> zone axis. The size of the in-phase octahedral tilted domains (bright contrast) varied from 10 to 40 nm across. The dark-contrast in the dark-field image corresponds to the regions with tilting along other directions or without the in-phase octahedral tilting at all. Fig. 7(b) is a dark-field image obtained using one of the γ reflections along a <011> zone axis. The size of the anti-phase octahedral tilted domains is ~ 10 nm across, which is smaller than those of the in-phase tilted one. The distributions of these two octahedral tilted domains are fairly uniform. It seems that these domains have not been influenced by the sintering process, which can cause the spatial preferential 1:1 cation ordering near grain boundaries found in many complex perovskite ceramics.[45,46,47] The well-developed anti-phase boundaries separating the macroscopic octahedral tilted domains, which can be seen in other complex perovskite titanate solid solutions,[48] are not observed. It indicates that the octahedral tilting occurred in BNKLBT-1.5 ceramic is confined in nano-scale only. Since there is a lack of macroscopic / long-range octahedral tilted domains, we are unable to determine the octahedral tilt system by tilting the specimen at different zone axes systematically.

**IV. Discussion**

The observed propeller-shaped hysteresis loops shown in Fig. 1 is not the characteristic of an antiferroelectric material because the forward switching field ($E_{AFE-FE}$) at which the antiferroelectric domains are aligned to become ferroelectric domains cannot be seen. In fact, the polarizations increases continuously and linearly without a steep increase until approaching to the saturation when the applied field increases. This is similar to the behavior of a relaxor. When reducing the electric field, the polarizations gradually decreases but a sudden drop was noted at 1 MV/m, at which the field-induced polar domains returned to their non-polar state. Therefore, this behavior is highly likely to be the field response of the mixture of spontaneous and field-induced polarizations existed in the sample. We exclude the possibility of antiferroelectricity because the corresponding crystallographic evidence is not found in the *in-situ* TEM study. Apparently, the slimness of the hysteresis loops is resulted by the weakened ferroelectric domains. At the high temperatures, the morphology of the ferroelectric domains and their walls, which are shown in Fig. 2, indicates the decrease of spontaneous polarizations. In addition, the SAED study reveals that the octahedral tilting also diminishes with increasing temperature. As a result, we deduce that the ferroelectric domains in BNKLBT-1.5 are highly influenced by the octahedral tilting.

In perovskites, the stabilities of lattice are often discussed in term of the tolerance factor,[49]

$t = \dfrac{(r_A + r_O)}{\sqrt{2}(r_B + r_O)}$ where $r_A$, $r_B$ and $r_O$ are the respective radii of the A-site, B-site and oxygen ions.

The value of t for an ideal and undistorted perovskite is unity. In general, a perovskite with t < 1

may contain octahedral tilt to stabilize the structure in which a small cation occupies the A-site.[37,39,50] In the present study, $r_A$ is substituted by the average ionic radii of the A-site cations $<r_A>$. In the calculation of $<r_A>$, it can be convincingly assumed that the cations are randomly distributed and no cation order is formed in the A-sites because TEM results show no long-range 1:1 cation order. The ionic radii of the individual and averaged ionic radii and tolerance factors for some Bi-based titanates are listed in Table I. The tolerance factor for BNKLBT-1.5 is 0.9847, implying the possibility of octahedral tilting. On the other hand, the values of t for the different compositions, which are possible to exist locally in BNKLBT-1.5, vary from 0.98 to 1.06. This indicates that various kinds or degrees of octahedral tilting can establish in the regions with different compositions. This proposal of the compositional dependence of octahedral titling is consistent to the dark-field TEM results showing the nano-sized octahedral tilted domains. The compositional dependent phase transition from the tilted to non-tilted states in BNKLBT-1.5 ceramics is analogous to that of many disordered perovskite relaxor-ferroelectric solid solutions,[51] in which macroscopic ferroelectric domains develop below the phase transition temperature though the chemical compositions of the cation sites are in a disordered state.

At low temperatures, various localized octahedral tilts compensate the lattice distortions introduced by different size of A-site cations and hence stabilize the macroscopic ferroelectric domains. When increasing temperature, the octahedral tilted domains disappear gradually and hence weaken the ferroelectric domains progressively, as seen in Fig. 2. We suggest that the non-tilted regions cannot well-stabilize the lattice in some compositions and therefore at high temperature the long-range polar order is vanished in the non-tilted regions. However, the macroscopic transition from the tilted to non-tilted state does not occur abruptly at $T_d$. The individual octahedral transformation depends on the local arrangement of the A-site cations. The local compositional variations give rise to different lattice distortions[24] and also transition temperatures.[9] As the results, the forbidden reflection and ferroelectric domains are still observed above $T_d$. On the other hand, the long-range antiferroelectric order in the ceramic is broken by the A-site cation disorder in the crystal structures, that is similar to the case of other complex perovskites,[27] in which the presence of either commensurate or incommensurate antiferroelectric domains highly depends on the degree of the cation order.

The peculiar propeller-shape hysteresis loop at high temperature can be explained as follows. The field response of the polar regions in BNKLBT-1.5 is the typical ferroelectric behavior that contributes to the $P_r$ at zero field and $E_c$. When the applied field increases, these polarizations are re-orientated. On the other hand, some non-polar regions start to develop by shifting the A-site cations and hence the neighboring octahedral may be tilted, or vice versa. This field-induced coupling is to stabilize to the lattice and also to establish the aligned polar state. The long-range polar order can then developed continuously until the saturation. It should be noted that not all the non-polar regions can become polar regions. Such transformation highly depends on the transition temperature of each nano-sized domain and also the strength of the applied field. When the applied field is decreased, the polarization decreases linearly until a critical field at 1 MV/m, at which the polarization drops suddenly. It is suggested that 1 MV/m is the minimum field to maintain the field-induced coupling and the polar states in BNKLBT-1.5. Below this threshold, the field-induced polar domains return to their original non-polar states and the polar domains dominates the polarization response of the ceramic.

## V. Conclusions

The change in domain morphology in the lead-free piezoelectric BNKLBT-1.5 ceramic was investigated by *in-situ* TEM. The search for the antiferroelectric domains was carried out. Neither commensurate nor incommensurate antiferroelectric domains are found at 25 °C to 180 °C. It is concluded that the depolarization in BNKLBT-1.5 is not originated from the ferroelectric-antiferroelectric transition. The depolarization is induced by the weakening of the macroscopic ferroelectric domains, which is associated with the reduction of octahedral tilting. Our TEM results reveal that ferroelectric domains and octahedral tilting do not vanish instantaneously at $T_d$ or a particular temperature. The macroscopic tilt to non-tilt transition is diffusive and the composition dependency of such transition is suggested. Hence, we suggest that the frequency dispersion in the dielectric response of BNKLBT-1.5 ceramic could also relate to this transition.[52]

In conclusion, octahedral tilting occurs in BNKLBT-1.5 ceramic and plays an important role on the temperature-dependent ferroelectric domain morphology, electrical and electromechanical properties. We suggest that octahedral tilting could occur in other BNT-based solid solutions which show the depolarization and peculiar propeller-shaped hysteresis loops at the vicinity of $T_d$. However, further investigations are needed to understand the configurations of the local octahedral tilting and their influence on the polar states in BNKLBT-1.5 ceramic, as well as in other BNT-based solid solutions.

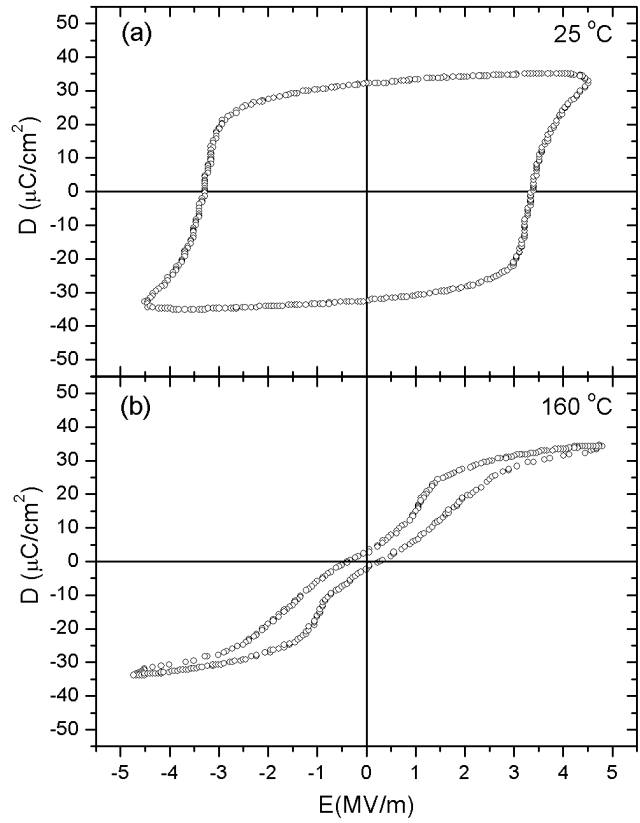

Fig. 1. Ferroelectric (D-E) hysteresis loops of BNKLBT-1.5 ceramic measured at (a) 25 °C and (b) 160 °C, respectively.

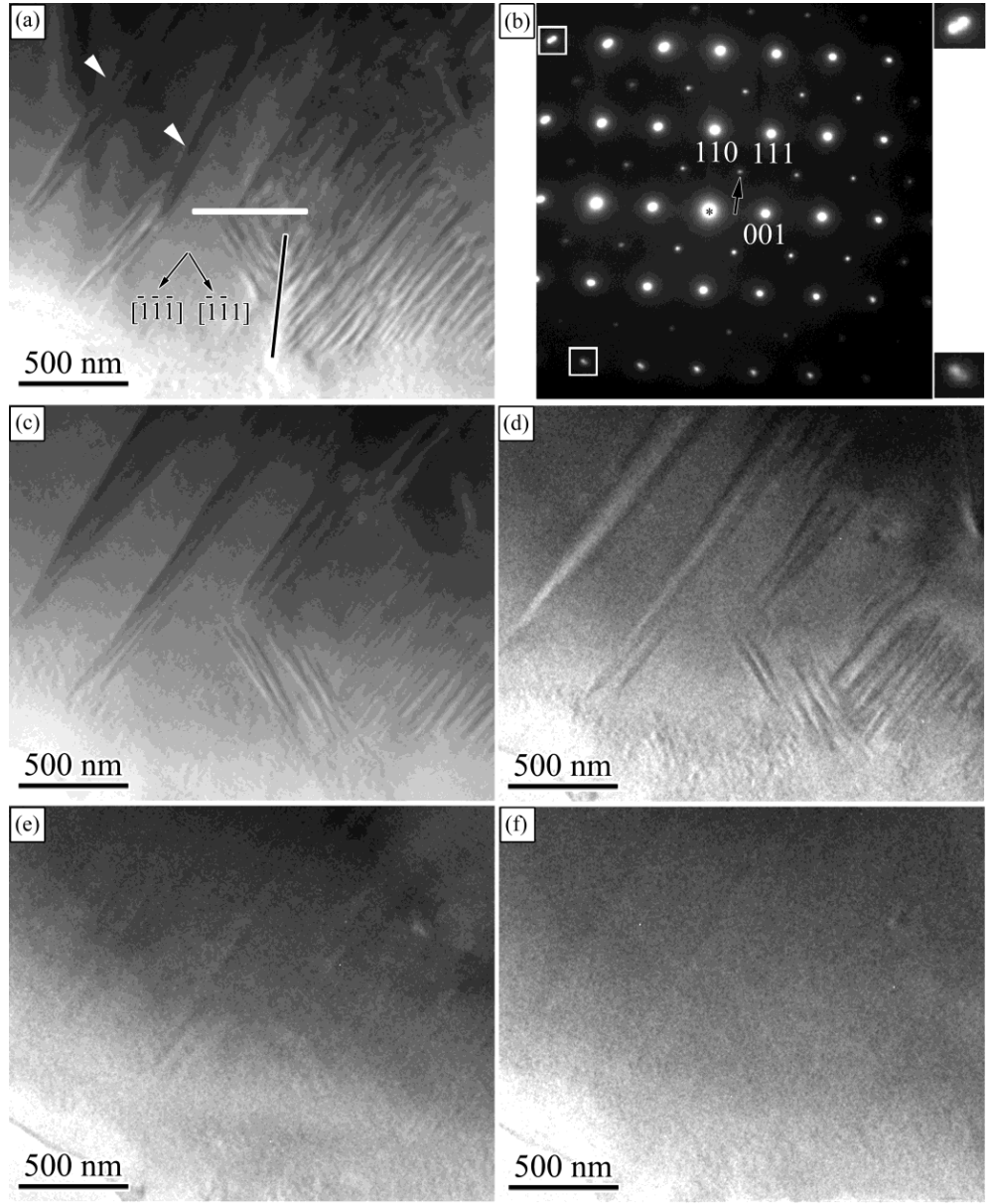

Fig. 2. Bright-field TEM images of the ferroelectric domains in the ceramic recorded at (a) 25 °C, (c) 100 °C, (d) 140 °C, (e) 160 °C and (f) 180 °C, respectively. The (101) and (011) domain walls are marked with a white and black line, respectively. (b) SAED pattern of the same region taken at 25 °C. A superstructure 1/2 1/2 1/2 reflection is indicated by an arrow. Insets: two reflections are enlarged to show the splitting of the spots.

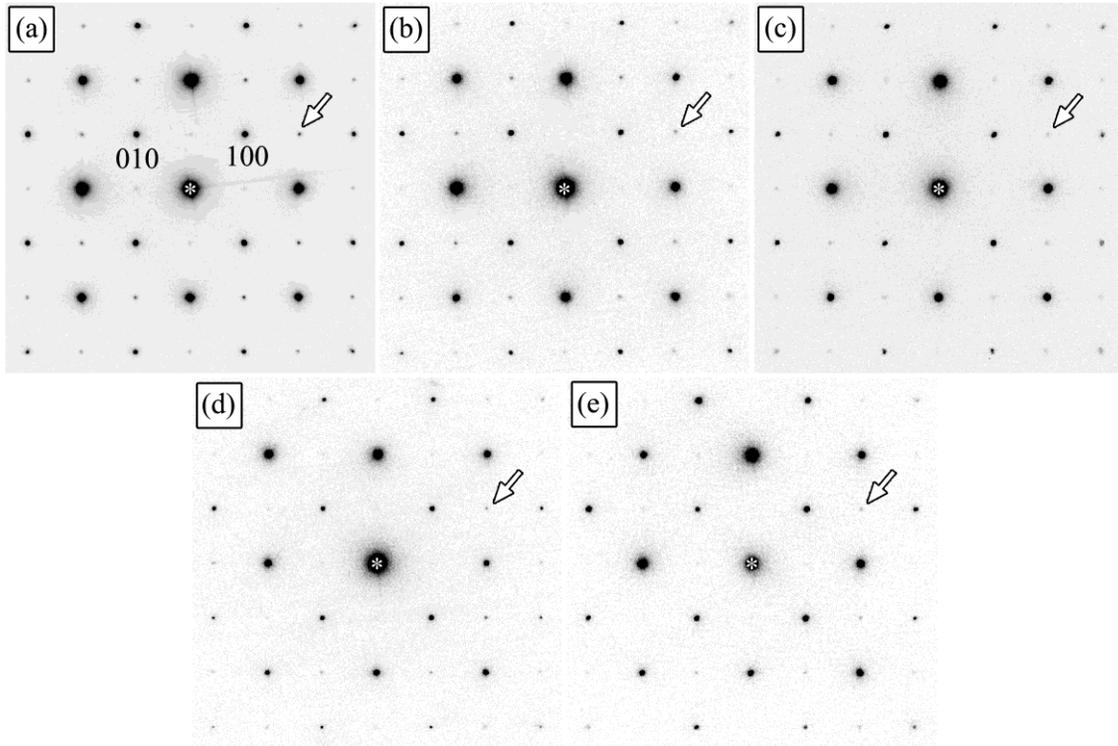

Fig. 3. SAED patterns recorded along a <001> zone axis of a grain of BNKLBT-1.5 ceramic at (a) 25 °C, (b) 100 °C, (c) 140 °C, (d) 160 °C and (e) 180 °C. The forbidden 3/2 1/2 0 reflections are marked by the white arrows. The diffraction patterns are shown as negatives in order to make the weak forbidden reflections clearer.

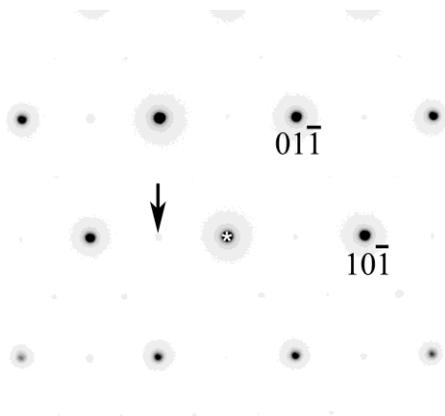

Fig. 4. SAED pattern recorded along a <111> zone axis of a grain of BNKLBT-1.5 ceramic at 25 °C. The γ reflections can be seen. The forbidden reflection -1/2 0 1/2 is indicated by an arrow.

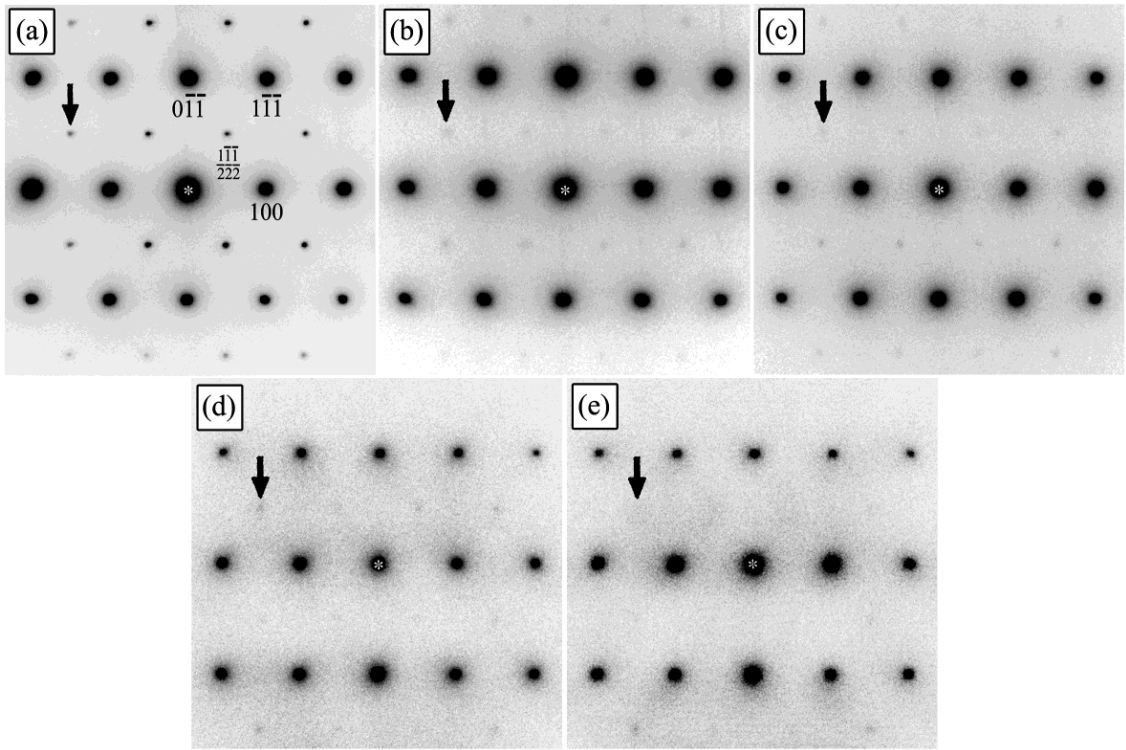

Fig. 5. SAED patterns recorded along a <011> zone axis of a grain of BNKLBT-1.5 ceramic at (a) 25 °C, (b) 100 °C, (c) 140 °C, (d) 160 °C and (e) 180 °C. The forbidden -3/2 -1/2 -1/2 reflections are indicated by the black arrows.

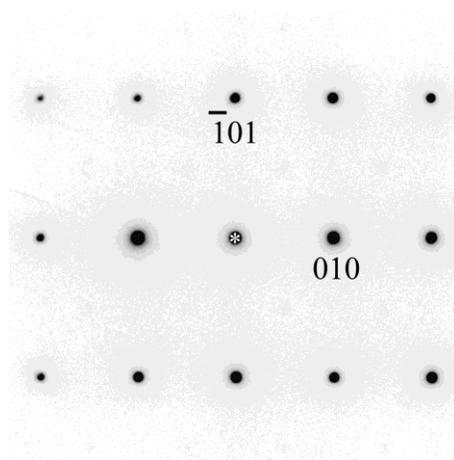

Fig. 6. SAED pattern recorded along a <011> zone axis of a grain of BNKLBT-1.5 ceramic at 25 °C. No forbidden reflection can be seen.

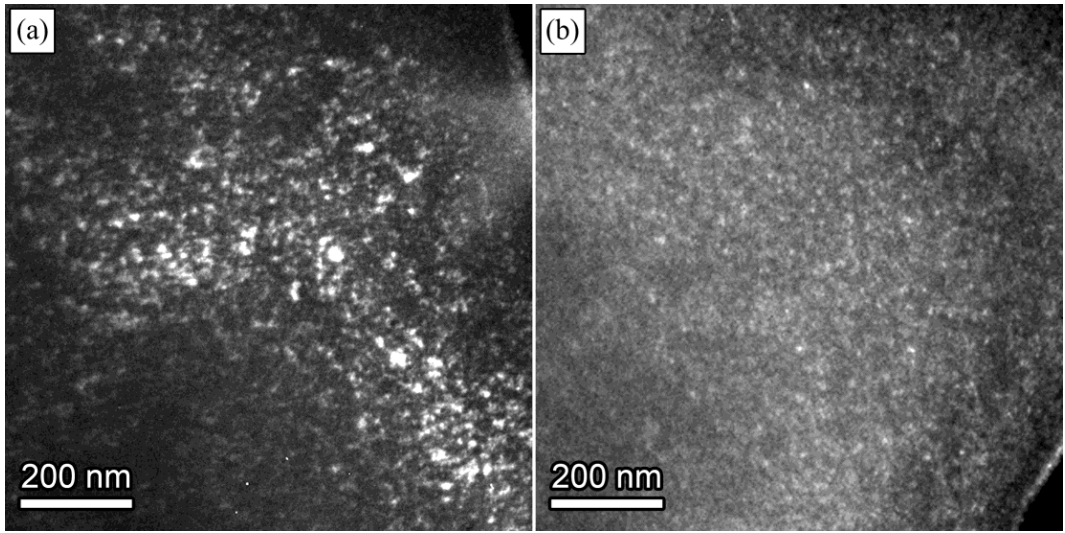

Fig. 7. Dark-field images of the forbidden (a) 3/2 1/2 0 and (b) 3/2 1/2 1/2 reflections recorded at room temperature. Only the titled regions, which scatter the electron into (3/2 1/2 0) or (3/2 1/2 1/2), contribute to the bright contrast in (a) or (b), respectively.

Table I. Average ionic radii of A-site cations and tolerance factors for Bi-based titanates and their solid solutions. The listed compositions may exist in BNKLBT-1.5 ceramic locally. The calculated values are based on the ionic radii listed in Shannon's table ($O^{2-}$ = 1.40 Å, $Ti^{4+}$ = 0.605 Å, $Na^+$=1.39 Å, $K^+$ = 1.64 Å and $Ba^{2+}$ = 1.61 Å).[53] The ionic radii of $Bi^{3+}$ and $Li^+$ for coordination number XII are obtained by extrapolation of their values for the lower coordination numbers, since they are not available in Ref. [53]. The extrapolated ionic radii of $Bi^{3+}$ and $Li^+$ are 1.36 Å and 1.24 Å, respectively.

| Compositions | Average ionic radii $\langle r_a \rangle$ (Å) | Tolerance factor |
|---|---|---|
| $Bi_{1/2}Na_{1/2}TiO_3$ (BNT) | 1.375 | 0.9787 |
| $Bi_{1/2}K_{1/2}TiO_3$ (BKT) | 1.50 | 1.0227 |
| $Bi_{1/2}Li_{1/2}TiO_3$ (BLT) | 1.30 | 0.9522 |
| $BaTiO_3$ (BT) | 1.61 | 1.0615 |
| 0.885BNT-0.05BKT-0.015BLT-0.05BT (BNKLBT-1.5) | 1.392 | 0.9847 |
| 0.95BNT-0.05BKT | 1.381 | 0.9808 |
| 0.985BNT-0.015BLT | 1.374 | 0.9783 |
| 0.95BNT-0.05BT | 1.387 | 0.9829 |
| 0.77BKT-0.23BLT | 1.454 | 1.0065 |
| 0.5BKT-0.5BT | 1.555 | 1.0421 |
| 0.23BLT-0.77BT | 1.539 | 1.0365 |
| 0.935BNT-0.05BKT-0.015BLT | 1.380 | 0.9804 |
| 0.9BNT-0.05BKT-0.05BT | 1.393 | 0.9850 |
| 0.435BKT-0.13BLT-0.435BT | 1.522 | 1.0305 |

Incommensuration in the Intermediate Phase Region of Lead Zirconate," *J. Am. Ceram. Soc.*, **78** [8] 2220-2224 (1995).

[29]Z. Xu, X. Dai and D. Viehland, "Incommensuration in La-modified Antiferroelectric Lead Zirconate Titanate Ceramics," *Appl. Phys. Lett.*, **65** [25] 3287-3289 (1994).

[30]G. C. Edwards, S. H. Choy, H. L. W. Chan, D. A. Scott, A. Batten, "Lead-free Transducer for Non-destructive Evaluation," *Appl. Phys. A*, **88** [1] 209-215 (2007).

[31]S. H. Choy, X. X. Wang, H. L. W. Chan and C. L. Choy, "Electromechanical and Ferroelectric Properties of $(Bi_{1/2}Na_{1/2})TiO_3$-$(Bi_{1/2}K_{1/2})TiO_3$-$(Bi_{1/2}Li_{1/2})TiO_3$-$BaTiO_3$ Lead-free Piezoelectric Ceramics for Accelerometer Application," *Appl. Phys. A*, **89** [3] 775-781 (2007).

[32]J. Ricote, R. W. Whatmore and D. J. Barber, "Studies of the Ferroelectric Domain Configuration and Polarization of Rhombohedral PZT Ceramics," *J. Phys.: Condens. Matter.*, **12** [3] 323-337 (2000).

[33] M. Verwerft, G. Tendeloo, J. Landuyt and S. Amelinckx, "Electron Microscopy of Domain-Structures," *Ferroelectrics*, **97**, 5-17 (1989).

[34]P. G. Lucuta and V. Teodorescu, "SEM, SAED, and TEM Investigations of Domain Structure in PZT Ceramics at Morphotropic Phase Boundary," *Appl. Phys. A*, **37** [4] 237-242 (1985).

[35]I. M. Reaney, E. L. Colla and N. Setter, "Dielectric and Structural Characteristics of Ba- and Sr-based Complex Perovskites as a Function of Tolerance Factor," *Jpn. J. Appl. Phys.*, **33** [7] 3984-3990 (1994).

[36]J. Ricote, D. L. Corker, R. W. Whatmore, S. A. Impey, A. M. Glazer, J. Dec and K. Roleder, "A TEM and Neutron Diffraction Study of the Local Structure in the Rhombohedral Phase of Lead Zirconate Titanate," *J. Phys.: Condens. Matter*, **10** [8] 1767-1786 (1998).

[37]A. M. Glazer, "The Classification of Tilted Octahedra in Perovskites," *Acta Crystallogr. B*, **28** [11]
24